\documentclass[twocolumn,showpacs,superscriptaddress]{revtex4}

\usepackage{amsmath,amssymb}
\usepackage{graphicx}
\usepackage{dcolumn}
\usepackage{bm}
\usepackage{epstopdf}
\usepackage{amstext}
\usepackage{soul}
\usepackage[usenames, dvipsnames]{color}

\begin{document}

\title{Deterministic quantum state transfer between remote atoms with photon-number superposition states}

\author{A. Gogyan }
\email{agogyan@gmail.com} \affiliation{Institute for Physical
Research, Armenian National Academy of Sciences, Ashtarak-2, 0203,
Armenia}
\author{Yu. Malakyan}
\email{yumal@ipr.sci.am}
\affiliation{Institute for Physical Research, Armenian National
Academy of Sciences, Ashtarak-2, 0203, Armenia}

\begin{abstract}
We propose a protocol for quantum networking based on deterministic quantum state transfer between distant memory nodes using photon-number superposition states (PNSS). In the suggested scheme, the quantum nodes are single atoms confined in high-finesse optical cavities linked by photonic channels. The quantum information written in a superposition of atomic Zeeman states of sending system is faithfully mapped through cavity-assisted Raman scattering onto PNSS of linearly polarized cavity photons. The photons travel to the receiving cavity, where they are coherently absorbed with unit probability creating the same superposition state of the second atom, thus ensuring high-fidelity transfer between distant nodes. We  develop this approach at first for photonic qubit and show that this superposition state is no less reliably protected against the propagation losses compared to the single-photon polarization states, whereas the limitation associated with the delivery of more than one photon does not affect the process fidelity. Then, by  preserving the advantages of qubits, we extend the developed technique to the case of state transfer by photonic qutrit, which evidently possesses more information capacity. This reliable and efficient scheme promises also a successful distribution of entanglement over long distances in quantum networks.
\end{abstract}

 \maketitle
\section{\protect\normalsize INTRODUCTION}

Quantum networks are the platform for the implementation of quantum communication and quantum information processing \cite{1}. They comprise of local network nodes, which are capable of efficient storing, processing and releasing photonic quantum information in a reversible fashion, and of quantum channels, which link the memory nodes by transmitting photonic states between them with high fidelity of the transfer. The original proposal for the quantum state transfer (QST) is based on the "single atom-optical cavity" system \cite{2}, where the reversible mapping of quantum states between light and matter is achieved due to the strong coupling of the interaction of single atoms and photons, while quantum channels, for example optical fibers, coherently exchange single photons between different nodes. The remarkable capability of this scheme to connect distant quantum objects, stimulated many efforts aimed at solving the main challenges in quantum information transfer between remote qubits. Specifically, new schemes have been developed for robust creation and storage of quantum entanglement over long distances \cite{A,B,C,D}, heralded storage of polarization states of single photons in single atoms \cite{E,kurz,F}, quantum error correction \cite{G}, implementation of quantum gates between distant atoms \cite{H}, single-photon generation \cite{I, J, K}, faithful and controllable matter-light quantum interfaces \cite{L, M} including those based on optomechanical transducers \cite{N}. Utilizing the global resources of this scheme, a new quantum protocol have been offered for a network connecting the geographically remote optical atomic clocks \cite{O}.  Inspired by the proposal of Ref.\cite{2}, where the atom-cavity system produces a time-symmetric photon wave packet to prevent the photon reflection from the second cavity mirror, a different approach was suggested in \cite{P} based on an adiabatic passage through a dark state of both cavities that reduces the cavity photon losses. The first implementation of the scheme in Ref.\cite{2} with direct, albeit probabilistic, connection of the states of two distant atoms was presented in the recent paper \cite{3}, which combines in one experiment the basic operations that have been previously realized separately with cavity QED including the interconversion of photonic and atomic quantum states \cite{4,5} and generation and storage of cavity photons in a single-atom quantum memory \cite{6}. However, the  efficiency of the QST currently available in this scheme is rather low to implement scalable quantum networks, as a desired result is obtained only after many unsuccessful attempts requiring much more time as compared to the realistic quantum memory lifetimes. These difficulties are typical also for the entanglement swapping in probabilistic protocols for quantum repeaters \cite{7}. For the full employment of quantum networks, new schemes with highly efficient emission and storage of cavity photons, as well as low-loss photonic links between the cavities are required. Recently, the deterministic transfer of microwave photons and entanglement generation between distant superconducting qubits have been demonstrated in a number of works \cite{8,9,10}, but the reliable exchange of optical photons between spatially separated nodes still remains a challenging task.

\begin{figure*}[t] \rotatebox{0}{\includegraphics*
[scale =0.7]{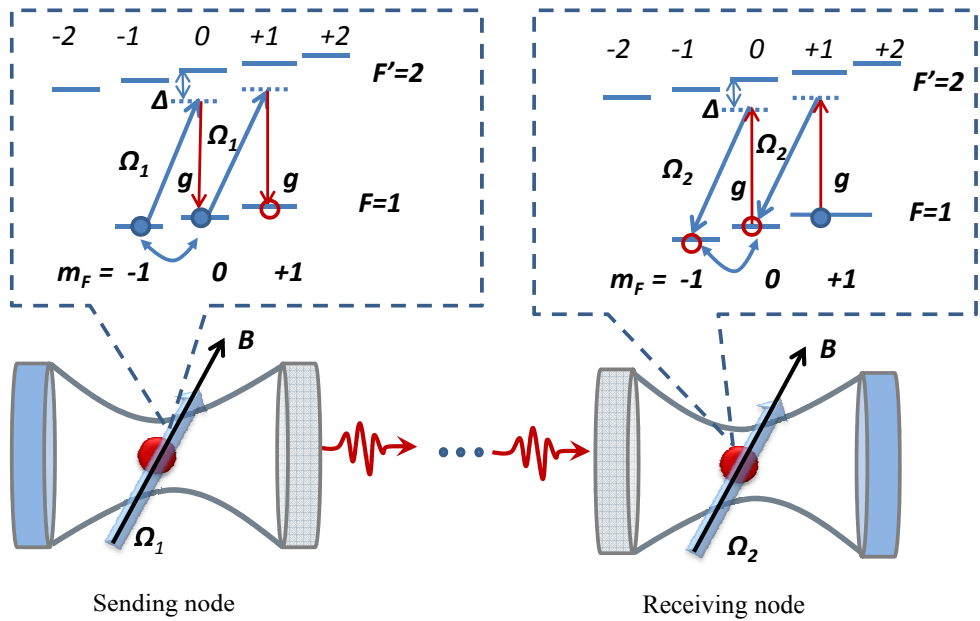}} \caption{(Color online) (a) Quantum state transfer between two single-atom network nodes. The insets show the atomic level structure and transitions in the laser fields $\Omega_{1,2}$ and in an external magnetic field B. The sequent photons (thin red lines) are generated in the left and absorbed in the right cavities. The initial populations of atomic states are shown by filled circles \label{fig:fig1}}
\end{figure*}

In this paper, we propose a quantum network protocol based on the deterministic generation and absorption of cavity photons in network nodes, which consist of single $F$-Zeeman-structured multilevel atoms, with $-F\le m_F\le +F$, confined in identical single-mode high-finesse optical cavities. In the sending node, an arbitrary quantum state is encoded in the superposition state of the ground Zeeman sublevels of the single atom, which strongly interacts with linearly polarized cavity mode and a circularly $\sigma^+$ polarized control $\Omega_1(t)$ laser field in a Raman configuration. The control field transfers the atomic population into the extreme Zeeman state $m_F=+1$ via Raman transition and, thus, completely converts the atomic superposition state into PNSS of linearly polarized cavity photons.  Note that the schemes enabling atomic ground-state Zeeman coherence to be mapped onto the cavity-mode field using the adiabatic passage have been studied in \cite{first,second}. In our case of Raman configuration the temporal profiles of the photons are determined by the shape of laser pulse. This process is deterministic due to the very high signal-to-noise ratio that is available in real conditions. We first consider the simple case of the atomic $F=1\rightarrow F'=2$ transition, when the atom is initially prepared in an arbitrary superposition of ground Zeeman states $m_F=-1$ and $m_F=0$ (Fig.\ref{fig:fig1}, left), where the photonic qubit as a superposition of one- and two-photon states is generated, which identically reproduces the form of atomic superposition state. An external magnetic field applied perpendicular to the cavity axis removes the Zeeman degeneracy in such a way as to fine-tune the linearly polarized cavity mode to the exact resonance with two-photon Raman-transitions, as well as to reduce the decoherence of the atomic superposition by ambient magnetic fields. The photons leave the cavity through one mirror, which is partially transparent, as a freely propagating wave packets, which are efficiently coupled to a low-loss single-mode optical fiber. The latter transmits the photons to the receiving system (Fig.\ref{fig:fig1}, right), where a single atom is prepared initially in the state $m_F=+1$. The atom interacts with a circularly $\sigma^+$ polarized $\Omega_2(t)$ laser field, which induces coherent absorption of incoming photons with unit probability, eventually settling down the receiving atom into the same superposition state as the original one in the sending system. The shape of the $\Omega_2(t)$ laser pulse is derived from the requirement to entirely exclude the photon leakage from the second cavity indicating that the quantum output field is zero at all times. We show that the absorption amplitudes are not sensitive to the photons' temporal shape due to integral dependence on the latter (see Sec.IIIB) that significantly alleviates the strict condition of high transfer efficiency requiring the photons of time-symmetric shapes \cite{2,11}.

Further we extend the developed technique to the case of the QST by the photonic qutrit, which is a superposition of the vacuum, one- and two-photon states and possesses greater information capacity, whereas the quantum connectivity between the nodes using this state is not less protected from the decoherence induced by the environmental effects than the photonic polarization qubit \cite{3}. With reduced photon losses in quantum channels, this deterministic protocol provides very fast and robust QST over large distances that notably mitigates the limitations on the quantum memory lifetime. Under these conditions the imperfection associated with the delivery of more than one photon does not reduce the final state fidelity.

This paper is organized as follows: In the next section, we present the interaction setup and derive the basic equations for the time evolution of the atomic state amplitudes and the cavity field, which describe the generation of PNSS of linearly polarized cavity photons in the sending system. We analyze the dynamics of atomic population and photon number distribution via numerical calculations, as well as we obtain the temporal profiles of emitted photons. Here we also estimate the error due to the spontaneous losses and discuss the main approximations ensuring the deterministic generation of PNSS. The analytic solutions for the coherent absorption of PNSS in the receiving system are found in Sec.III, where the final superposition state of the second atom is demonstrated. Here we also present the implementation of the QST via photonic qutrit. We discuss also the advantages and imperfections of our protocol in the Sec.IV and summarize the results in Sec.V.

\section{\protect\normalsize PNSS GENERATION IN THE SENDING NODE}

\subsection{Model}
The present mechanism for producing PNSS is based on our earlier proposed method of deterministic generation of a stream of multiphoton pulses in a single-atom-single-mode cavity QED system \cite{12,13}. A multi-level atom or ion is trapped in a one-mode high-finesse optical cavity and interacts with a $\sigma^+$-polarized laser field $\Omega_1$ (Fig.\ref{fig:fig1}a) on the multi-level chain, for instance, on the transition $5S_{1/2}(F = 1) \rightarrow 5P_{3/2}(F' = 2)$ of the $^{87}$Rb atom, where the state $5P_{3/2}(F' = 2)$ is well isolated from other hyperfine levels. The laser field couples the ground state $F=1$ with magnetic quantum number $m_F = -F,\cdots,F-1$ and the excited state $|F' =2, m_{F'} = m_F+1\rangle_1$ and creates a linearly polarized cavity-mode Stokes-photon (shown in Fig.\ref{fig:fig1} by red thin lines) on the transitions $|F' =2, m_{F'} = m_F+1\rangle_1 \rightarrow |F =1, m_F+1\rangle_1$, thus transferring the atom into the next Zeeman sublevel with $m_F+1$.  The one-photon blue detuning of the laser field, which is the same for both atoms $\Delta_1$=$\Delta_2$=$\Delta$, is taken very small compared to the upper level hyperfine splitting, but at the same time it is much larger than the cavity decay rate $k$, the natural spontaneous decay rate $\gamma_{\text{sp}}$ of the atom and the Rabi and Larmor frequencies: $|\Delta| \gg k, \gamma_{\text{sp}}, \Omega_{1}, \Delta_B^{(F, F')}$, where $\Delta_B ^{(F,F')}=g_L ^{(F,F')}\mu_B B/\hbar$ is the Zeeman splitting of the ground and excited states in the magnetic field $B$, with $g_L^{(F, F')}$ the Land\'e factor and $\mu_B$ the Bohr magneton. This condition allows one to suppress the off-resonant excitation to nearby hyperfine state $F'=1$ and, hence, to neglect the spontaneous losses from upper levels and dephasing effects induced by other excited states. In this far off-resonant case and for slowly varying laser fields: $d\Omega_1 / dt \ll \Delta\Omega_1$, one can adiabatically eliminate the upper atomic states that leads to the effective Raman atom-photon coupling
\begin{equation}
G_1=g\Omega_1/\Delta,\label{eq:1}
\end{equation}
which can be made much weaker than the cavity field decay: $G_1\ll k$. This ensures that the generated photons leave the cavity without being reabsorbed by the atom, resulting in the deterministic emission of photons. For simplicity, we assume here that the dipole matrix element $\mu _{F,F'}$ of the $F\leftrightarrow F'$ transition does not depend on the magnetic quantum numbers. This actual dependence can be easily taken into account in the numerical simulations \cite{13}.

The external magnetic field taken as the quantization axis is applied orthogonal to the cavity axis and parallel to the laser pulse propagation. As we mentioned in the introduction, it is chosen strong enough to preclude the generation of $\sigma^+$ and $\sigma^-$ polarized cavity photons due to the off-resonance interaction with the atom, which is the case if $\Delta_B^{(F)}\gg k$. This clearly increases the probability of photon production with linear polarization unlike the model considered in Ref.\cite{3}, where the generation of cavity photons with undesired polarization cannot be suppressed.

We describe the dynamics of the system in the Heisenberg picture. The pumping laser field is given by
\begin{equation}
E_1(t)={\mathcal E}_1f_1^{1/2}(t)\exp (-i\omega t)+h.c.,\label{eq:2}
\end{equation}
where $f_1(t)$ features its temporal profile of duration $T_1$ and ${\mathcal E}_1$ is the peak amplitude of the field with one-photon detuning $\Delta=\omega - \omega_{FF'}$ and the peak Rabi frequency $\Omega_1=\mu_{F,F'}{\mathcal E}_1/\hbar$.

After eliminating the upper states by virtue of large one-photon detuning, the effective interaction Hamiltonian in the RWA takes the form
\begin{align}
H&=\hbar \Biggl[\sum_{m_F=-F}^{F}\biggl ( \frac{g^2}{\Delta}a_1^{\dag}a_1+f_1(t)\frac{\Omega_1^2}{\Delta}\biggr )\sigma_{m_F,m_F} \nonumber \\
&+ G_1 f_1^{1/2}(t) \sum_{m_F=-F}^{F-1}\biggl (a_1^{\dag}(t) \sigma_{m_F+1,m_F}(t)+h.c.\biggr )\Biggr],\label{eq:3}
\end{align}
where $\sigma_{i,j}(t) = |i \rangle_{11}\langle j |$ and $a_1(t)$, $a_1^\dag(t)$ are correspondingly the atomic and photonic mode operators in the first cavity. The first and second terms in Eq.\eqref{eq:3} describe the Stark shifts of atomic ground states induced by the cavity field and laser pulse $\Omega_1$, respectively. Under the adopted approximation of equal dipole moments, the Stark shifts induced by the laser field have no influence on the photon  generation, since their difference between two neighboring atomic ground states is zero for all transitions $m_F  \rightarrow m_F + 1$. On the other hand, the Stark shift $g^2/\Delta$ can be included into the cavity mode frequency: $\omega_c \rightarrow \omega_c-g^2/\Delta$. Below we use these simplifications to get the analytic solution revealing the main features of the system considered. In real atoms, where the dependence of $g$ and $\Omega_1$ on Clebsch-Gordan coefficients leads to non-vanishing Stark-shift difference, the latter can be made negligibly small as compared to the cavity decay rate $k$ by appropriately choosing the system parameters \cite{13}.

\subsection{Dynamics of atomic populations}

The production of PNSS from the first cavity starts by preparing the single atom in a superposition of the states $|F=1,m_F=-1\rangle_1$ and $|F=1,m_F=0\rangle_1$, denoted below as $|-1\rangle_1$ and $|0\rangle_1$, respectively, with normalized population amplitudes $c_{-1,0}: |c_{-1}|^2+|c_0|^2=1$, while the cavity mode is in the vacuum state. Then, the initial state of the sending system can be represented as
\begin{equation}
|\Psi_{\text{1,in}}\rangle=\biggl (c_{-1} |-1\rangle_1 + c_0 |0\rangle_1\biggr ) |0\rangle_{1c},\label{eq:4}
\end{equation}
where $|0\rangle_c$ refers to the state of cavity with zero photon number.

The applied $\Omega_1$ laser field transfers the atom to the final state $|F=1,m_F=+1\rangle_1=|1\rangle_1$ in two steps (if $c_{-1}\neq 0$), each time generating via Raman process a linearly polarized Stokes photon (Fig.\ref{fig:fig1},left).

The equations for the Zeeman sublevel populations $\langle\sigma_{m_F}(t)\rangle \equiv \langle\sigma_{m_F,m_F}(t)\rangle$ and ground-state coherence $\langle \sigma_{m_F,m_F+i} (t) \rangle=\langle \sigma^*_{m_F+i,m_F} (t) \rangle, i\neq 0,$ were derived in our previous papers \cite{12,13} from the master equation for the whole density matrix of the system.  The Hamiltonian \eqref{eq:3} and the input-output relation $a_{1,\text{out}}(t)-a_{1,\text{in}}(t)=\sqrt{k}a_1(t)$ \cite{14} for the photonic operators were used in the adiabatic limit $kT_1\gg 1$, where the cavity mode operator $a_1(t)$ is expressed through the atomic operators as
\begin{equation}
a_1(t)=-\frac{2i G_1}{k} f_1^{1/2}(t)\sum_{m_F=-F}^{F-1}\sigma_{m_F+1,m_F}(t) - \frac{2}{\sqrt{k}}a_{1,\text{in}}(t).\label{eq:5}
\end{equation}
In the sending node, the input field $a_{1,\text{in}}(t)$ is in the vacuum state $\langle a^{\dag}_{1,\text{in}}(t) a_{1,\text{in}}(t)\rangle = 0$ and will be ignored in further calculations.

In general, the atomic equations contain all relaxation processes including the optical pumping from $|F,m_F\rangle_1$ into states $|F,m_{F}+i\rangle_1, i = 1,2$, and the losses of atomic populations due to the decay from the upper atomic states into the states outside of the system. However, the deterministic production of cavity photons requires that the total spontaneous loss is negligible, which is achieved, if $\alpha_1f_1(t)\gg\Gamma_1(t)$, where $\alpha_1=4G_1^2/k$ is the cavity photon generation rate and $\Gamma_1(t)=\frac{\Omega_1^2}{\Delta^2}f_1(t)\gamma_{\text{sp}}$ is the total spontaneous decay rate induced by the $\Omega_1$ pump pulse. This defines the signal-to-noise ratio
\begin{equation}
R_{\text{sn}} = \frac{\alpha_1 f_1(t)}{\Gamma_{1}(t)} = \frac{4g^2}{ k\gamma_{\text{sp}}} \gg 1.\label{eq:6}
\end{equation}
This condition is clearly fulfilled in high-finesse optical cavity with $g > k, \gamma_{\text{sp}}$.

With this approximation the atomic equations are greatly simplified and for $F=1$ take the form
\begin{widetext}
\begin{subequations}
\begin{eqnarray}
& & \frac{d\langle \sigma_{m_F} (t) \rangle}{ dt}=\alpha_1f_1(t)[\langle \sigma_{m_F-1}(t) \rangle \theta(m_F)- \langle \sigma_{m_F}(t) \rangle \theta(-m_F)],\label{eq:7a}\\
& & \frac{d\langle \sigma_{m_F,m'_F} (t) \rangle }{ dt}= - \frac{1}{2}\alpha_1f_1(t)\biggl (\langle \sigma_{m_F,m'_F}(t)\rangle [\theta(-m_F)+\theta(-m'_F)] -2 \langle \sigma_{m_F-1,m'_F-1}(t)\rangle \theta(m_F)\theta(m'_F)\biggr ),\label{eq:7b}
\end{eqnarray}\label{eq:7}
\end{subequations}
\end{widetext}
where $\theta(x)$ is the Heavyside step function with $|m_F,m'_F|\leq F$ and $m'_F\neq m_F$.
One can easily check from Eq.\eqref{eq:7a} that the total population of atomic ground state is conserved:
\begin{equation}
\sum_{m_F=-F}^F\langle \sigma_{m_F} (t)\rangle  = 1.\label{eq:8}
\end{equation}

Equations \eqref{eq:7} are solved subjected to the initial conditions $\langle \sigma_{i}(-\infty)\rangle =|c_i|^2, i=-1,0$ and $\langle \sigma_{-1, 0}(-\infty)\rangle =c^*_{-1}c_0$, $\langle \sigma_{0,1}(-\infty)\rangle =\langle \sigma_{-1,1}(-\infty)\rangle=0$, which gives
\begin{subequations}
\begin{eqnarray}
& & \langle \sigma_{-1}(t)\rangle=|c_{-1}|^2 e^{-\vartheta(t)},\label{eq:9a}\\
& & \langle \sigma_0(t)\rangle=[|c_0|^2+|c_{-1}|^2 \vartheta(t)]e^{-\vartheta(t)},\label{eq:9b}\\
& & \langle \sigma_1(t)\rangle=1-(1+|c_{-1}|^2 \vartheta(t))e^{-\vartheta(t)},\label{eq:9c}\\
& & \langle \sigma_{-1,0}(t)\rangle=c^*_{-1}c_0 e^{-\vartheta(t)},\label{eq:9d}\\
& & \langle \sigma_{0,1}(t)\rangle=2c^*_{-1}c_0 (e^{-\vartheta(t)/2}-e^{-\vartheta(t)}),\label{eq:9e}\\
& & \langle \sigma_{-1,1}(t)\rangle\equiv 0,\label{eq:9f}
\end{eqnarray}\label{eq:9}
\end{subequations}
where the new variable
\begin{equation}
\vartheta(t) = \alpha_1\int \limits_{-\infty}^tf_1(t')dt'\label{eq:10}
\end{equation}
is proportional to the pump energy confined in the $(-\infty,t]$ area of the pulse.

\begin{figure}[h] \rotatebox{0}{\includegraphics*
[scale =0.7]{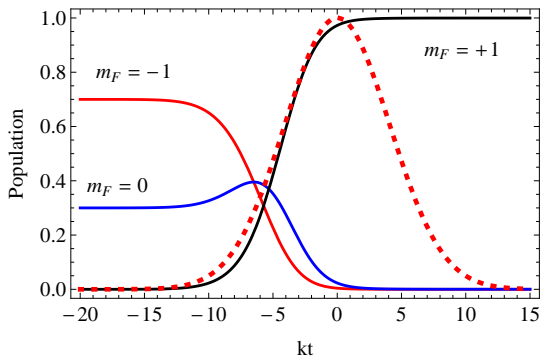}} \caption{(Color online) Populations of the ground Zeeman states $|m_F=-1\rangle_1$ (red),$|m_F=0\rangle_1$ (blue) and $|m_F=+1\rangle_1$ (black). The Gaussian laser pulse of FWHM duration 0.5~$\mu$s is shown by the dotted line.
The atom is initially prepared in the states $|F=1,m_F=-1\rangle_1$ and $|F=1,m_F=0\rangle_1$ with populations $|c_{-1}|^2=0.7$ and $|c_0|^2=0.3$, respectively. See the text for other parameters.
\label{fig:fig2}}
\end{figure}

For a Gaussian laser pulse of profile $f_1(t)=e^{-(t/T_1)^2}$ with duration $T_1=0.3~\mu$s, the atomic level populations are shown in Fig.\ref{fig:fig2}, where we have used the realistic parameters $(g, k, \gamma_{\text{sp}}, \Omega_1, \Delta^{F'}_B, \Delta) = 2\pi\times(12, 3, 5.87, 10, 15, 100)$~MHz \cite{15,16,17}, for which the signal-to-noise ratio in Eq.\eqref{eq:6} is $R_{\text{sn}} \sim 30$, thus justifying the approximation on neglecting relaxation processes. A magnetic field of $15$~G well resolves the Raman resonances, thus precluding the generation of circularly polarized cavity photons. There is a significant increase in the population of the state $|F=1, m_F=0\rangle_1 $ in the vicinity of $kt\sim-6$, which results from the population motion from the state $|F=1, m_F=-1\rangle_1$, while emitting the first photon. Yet, the both states are emptied almost simultaneously, which occurs already at the leading edge of the laser pulse. As shown in the next Section, this ensures simultaneous absorption of two photons, as a result the receiving atom settles in the desired superposition state.

After emission of photons, the sending atom passes to the final state $|F=1, m_F=+1\rangle_1$ (black line in Fig.\ref{fig:fig2}), where it continues to interact with the laser field on the transition $5S_{1/2}(F=1,m_F=+1) \rightarrow 5P_{3/2}(F'=2,m_{F'}=+2)$, neither emitting a cavity photon nor decaying into the outside state $5S_{1/2}(F=2)$ owing to the small value of $\Gamma_1$. Therefore, if the laser intensity rapidly tends to zero, the atom is left in the state $|F=1,m_F=+1\rangle_1$ and now can act as a receiving atom.

\subsection{Production of the output field state}
Here we analyze the output state of quantum field using Eqs.\eqref{eq:9}. The basis states of the sending system and outgoing photons are the atomic ground Zeeman states $|m_F\rangle$, vacuum state of the cavity mode $|0\rangle_{1c}$, and photonic states $|j\rangle$ with free propagating $j$ photons of frequency $\omega_c$, which are coupled to a single-mode optical fiber. The state $|j=2\rangle$ is not a $j$-Fock state, because two single-photon wave packets are in different temporal modes with wave functions $\Phi_i(t), i=1,2,$, which are found in Eqs.\eqref{eq:19}. The total state $|\Psi_1(t)\rangle$ can be then expanded in this basis as an entangled state between the atom and the outgoing light in the form
\begin{equation}
|\Psi_1(t)\rangle = \biggl (\sum_{m_F=-1}^{1}\sum_{j=0}^{m_F+1}\beta_{m_F,j}(t)|m_F\rangle_1|j\rangle\biggr )|0\rangle_{1c},\label{eq:11}
\end{equation}
where the coefficients $\beta_{m_F,j}(t)$ are normalized complex-valued amplitudes of the probabilities that at time $t$ the atom is in the state $|m_F\rangle_1$ and the number of photons in the resulting field is $j$. At $t\rightarrow -\infty$ the state $|\Psi_1(t)\rangle$ coincides with the initial state $|\Psi_{1,\text{in}}\rangle$ defined in Eq.\eqref{eq:4}. From this comparison, we find the initial values of $\beta_{-1,0}(-\infty)=c_{-1}$ and $\beta_{0,0}(-\infty)=c_{0}$, while the remaining coefficients are initially zero.

Our goal is to show that long after the interaction with the atom the state $|\Psi_1(t)\rangle$ is reduced to the final form
\begin{equation}
|\Psi_{1,\text{fin}}\rangle = |\Psi_1(\infty)\rangle = |+1\rangle_1|0\rangle_{1c} \otimes |\Psi_{\text {PNSS}}(t)\rangle,\label{eq:12}
\end{equation}
where the atom occupies the state $|m_F=+1\rangle$, while the output state of quantum field is the PNSS of two- and one-photon states
\begin{equation}
|\Psi_{\text {PNSS}}(t)\rangle=c_{-1}|1_{\Phi_1},1_{\Phi_2}\rangle + c_0 |1_{\Phi_1}\rangle.\label{eq:13}
\end{equation}
The coefficients $\beta_{m_F,j}(t)$ are connected to the populations of atomic states by the relation
\begin{equation}
\langle\sigma_{m_F}(t)\rangle = Tr[\rho_1(t)\sigma_{m_F}]=\sum_{j=0}^{m_F+1}|\beta_{m_F,j}(t)|^2,\label{eq:14}
\end{equation}
where $\rho_1(t)=|\Psi_1(t)\rangle \langle \Psi_1(t)|$.
Using Eqs.\eqref{eq:9}, $\beta_{m_F,j}(t)$ are found to be
\begin{subequations}
\begin{eqnarray}
& & \beta_{m_F=-1,j=0}(t)=c_{-1} e^{-\vartheta(t)/2},\label{eq:15a}\\
& & \beta_{m_F=0,j=0}(t)=c_0 e^{-\vartheta(t)/2},\label{eq:15b}\\
& & \beta_{m_F=0,j=1}(t)=c_{-1}[\vartheta(t) e^{-\vartheta(t)}]^{1/2},\label{eq:15c}\\
& & \beta_{m_F=1,j=0}(t)=0,\label{eq:15d} \\
& & \beta_{m_F=1,j=1}(t)=c_0 [1-e^{-\vartheta(t)}]^{1/2},\label{eq:15e}\\
& & \beta_{m_F=1,j=2}(t)=c_{-1}[1-(1+\vartheta(t))e^{-\vartheta(t)}]^{1/2},\label{eq:15f}
\end{eqnarray}\label{eq:15}
\end{subequations}
showing that at large times, when $\vartheta(t)e^{-\vartheta(t)}\ll1$, which is achieved at $kt\sim 3$, only two amplitudes stay on: $\beta_{m_F=1,j=1}(t\rightarrow\infty)=c_0$ and $\beta_{m_F=1,j=2}(t\rightarrow\infty)=c_{-1}$. Consequently, the state $|\Psi_1(t)\rangle$ is asymptotically transformed into $|\Psi_{1,\text{fin}}\rangle$ indicating that the quantum information encoded initially in the atomic superposition state $|\Psi_{1,\text{in}}\rangle$, Eq.\eqref{eq:4}, is completely converted into PNSS given by Eq.\eqref{eq:13}. Here it was assumed that no dephasing of photonic states occurs in the cavity mirror and communication channel.

\begin{figure}[h] \rotatebox{0}{\includegraphics*
[scale =0.7]{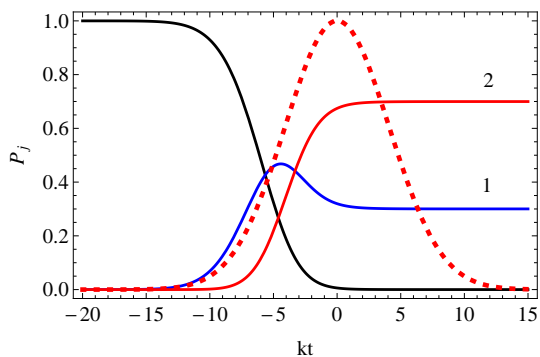}} \caption{(Color online) Dynamics of the photon number (indicated on the curves) distribution in the generated light for the same parameters as in Fig.\ref{fig:fig2}.
\label{fig:fig3}}
\end{figure}

Similarly, the photon distributions $P_j(t)$ in the resulting field, that is the population of the photonic state $|j\rangle$, is defined as the probability that the output field contains $j$ photons, no matter what state the atom is in, and, thereby, is given by the sum of $|\beta_{m_F,j}(t)|^2$ over all possible $m_F$
\begin{eqnarray} \label{solution}
P_{j}(t)= \sum_{m_F=-1}^1|\beta_{m_F,j}(t)|^2, \quad j=0,1,2. \label{eq:16}
\end{eqnarray}
$P_j(t)$ are calculated by means of Eqs.\eqref{eq:15} and are shown in Fig.\ref{fig:fig3} for the same parameters as in Fig.\ref{fig:fig2}. It is apparent that, as time increases, the states with $j=1,2$ remain populated as $|c_0|^2$ and $|c_{-1}|^2$, respectively, indicating the deterministic generation of PNSS.

\begin{figure}[h] \rotatebox{0}{\includegraphics*
[scale =0.7]{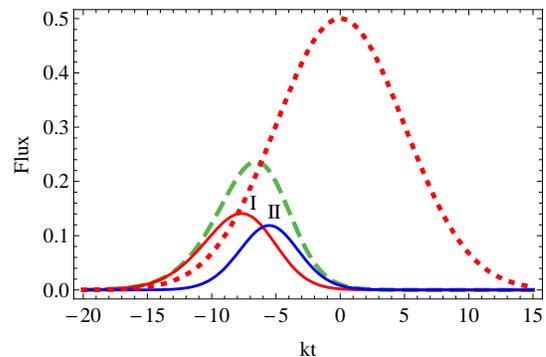}} \caption{(Color online) Partial fluxes of the first and the second photons $dn_j/dt,j=\text{I,II}$ (indicated on the curves) for the same conditions as in Fig.\ref{fig:fig2}. The total flux is shown by the green line. The laser field profile is decreased by factor of two.}
\label{fig:fig4}
\end{figure}

Now we find the flux of produced photons, which is needed to calculate the wave functions of emitted photons. In units of photons per unit time, the flux of the outgoing photons is defined by
\begin{equation}
\frac{dn_{\text{out}}}{dt}(t) = \langle a^{\dag}_{1,\text{out}}(t)a_{1,\text{out}}(t)\rangle,\label{eq:17}
\end{equation}
which determines the shape of emitted intensity. Here $n_{\text{out}}(t)$ is the mean photon number in the time interval $(-\infty,t]$ of the output field $a_{1,\text{out}}(t)$ from the first cavity, which is obtained from input-output relation and Eq.\eqref{eq:5}. Then, using Eq.\eqref{eq:8}, we have
\begin{equation}
\frac{dn_{\text{out}}}{dt}(t)=\alpha_1f_1(t)\sum_{m_F=-1}^{0}\langle \sigma_{m_F}(t)\rangle=\alpha_1f_1(t)[1-\langle \sigma_1(t)\rangle].\label{eq:18}
\end{equation}
The total flux is obviously the sum of the fluxes or intensities of the first and second photon pulses, which are readily separated in Eq.\eqref{eq:18} by substituting $\langle \sigma_1(t)\rangle$ from Eq.\eqref{eq:9c} that yields
\begin{subequations}
\begin{eqnarray}
\frac{dn_{\text{I}}}{dt}(t) &=& |\Phi_1(t)|^2= \alpha_1f_1(t)e^{-\vartheta(t)},\label{eq:19a}\\
\frac{dn_{\text{II}}}{dt}(t) &=& |c_{-1}|^2|\Phi_2(t)|^2 \nonumber\\
&=& |c_{-1}|^2\alpha_1f_1(t)\vartheta(t)e^{-\vartheta(t)},\label{eq:19b}
\end{eqnarray}\label{eq:19}
\end{subequations}
where the wave functions of the photons $\Phi_{1,2}(t)$ are considered to be real taking into account that they have the same phase as the control field $\Omega_1$, which can be taken zero without loss of generality. The partial fluxes corresponding to the preceding dynamics shown in Fig.\ref{fig:fig3} are displayed in Fig.\ref{fig:fig4}. The first photon intensity does not  evidently depend  on the population distribution between atomic states, while the second photon is emitted only if $c_{-1}\neq 0$. In both cases, the photon temporal profile is easily controlled being proportional to the $\Omega_1$ laser pulse shape.

The mean photon number is obtained by integrating Eq.\eqref{eq:18} yielding to
\begin{equation}
n_{\text{out}}(t)=(1+|c_{-1}|^2)(1-e^{-\vartheta(t)})-|c_{-1}|^2\vartheta(t)e^{-\vartheta(t)},\label{eq:20}
\end{equation}
which coincides with $n_{\text {total}}=\sum_{j=1}^{2} jP_{j}(t)$, as expected. This equation shows that for small $\vartheta(t)$, the generation of one photon dominates, where $n_{\text{out}}(t)=\vartheta(t)$  increases proportionally to the pump energy confined in that time interval, while two-photon emission is manifested at larger times, as it grows quadratically in $\vartheta(t)$ as $|c_{-1}|^2\vartheta^2(t)/2$.

Of particular importance is the question of joint generation of photons in the state $|1_{\Phi_1},1_{\Phi_2}\rangle$, which can be revealed by two photon correlation function at zero time delay
\begin{equation}
g^{(2)}(t) = \frac{\langle a_{\text{out}}^{\dagger 2}(t)a_{\text{out}}^2(t)\rangle}{\langle a_{\text{out}}^\dagger(t)a_{\text{out}}(t)\rangle^2}.\label{eq:21}
\end{equation}

\begin{figure}[h] \rotatebox{0}{\includegraphics*
[scale =0.7]{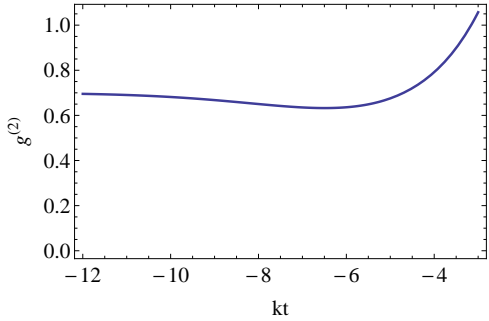}} \caption{(Color online) Second-order correlation function of generated photons at zero time delay for the same parameters as in Fig.\ref{fig:fig2}}
\label{fig:fig5}
\end{figure}

Here $\langle a_{\text{out}}^\dagger(t)a_{\text{out}}(t)\rangle^2$ gives the probability of emitting two single photons independently, while $\langle a_{\text{out}}^{\dagger2}(t) a_{\text{out}}^2(t)\rangle$ represents the probability of the generation of paired photons. In Fig.\ref{fig:fig5} the photon correlation displays a small antibunching in the main region $-10\leq kt\leq -3$ of the photon generation, indicating the predominance of the generation of uncorrelated photons compared to joint generation of photons. Moreover, the requirement that the intensity of the emitted light calculated in terms of PNSS coincides with Eq.\eqref{eq:19}
\begin{align}
& &\langle \Psi_{\text {PNSS}}(t)|a_{\text{out}}^{\dagger}(t) a_{\text{out}}(t)|\Psi_{\text {PNSS}}(t)\rangle \nonumber \\
& &=|\Phi_1(t)|^2+|c_{-1}|^2|\Phi_2(t)|^2\label{eq:22}
\end{align}
is satisfied, if the states $|1_{\Phi_1}\rangle$ and $|1_{\Phi_2}\rangle$ are orthogonal: $\langle 1_{\Phi_1}|1_{\Phi_2}\rangle=0$. In this case, the operator $a_{1,\text{out}}(t)$ can be represented as \cite{20}
\begin{equation}
a_{1,\text{out}}(t) = \sum_i\Phi_i(t)\hat b_i, \label{eq:23}
\end{equation}
where the independent annihilation operators $\hat b_i$ destroy the single-photon states in the usual way by operation $\hat b_i|1_{\Phi_i}\rangle=|0_{\Phi_i}\rangle$ and have the standard boson commutation relations $[\hat b_i,\hat b^\dagger_j]=\delta_{ij}$. 

\section{\protect\normalsize PNSS STORAGE IN THE RECEIVING NODE}
In the receiving node, the atom with the same level configuration is initially prepared in the state $|F=1,m_F=1\rangle_2$ and the cavity field is in the vacuum. Assuming that the losses in the communication channel are effectively switched off, the PNSS generated from the sending cavity constitutes the input field for the receiving one, thus the initial state of receiving system is given by
\begin{equation}
|\Psi_{2,\text{in}}\rangle = |+1\rangle_2 |0\rangle_{2c}\biggl (c_{-1}|1_{\Phi_1},1_{\Phi_2}\rangle + c_0 |1_{\Phi_1}\rangle\biggr ).\label{eq:24}
\end{equation}

The atom interacts with the $\sigma^+$ polarized laser field of frequency $\omega$ and the peak Rabi frequency $\Omega_2$, which can be different from $\Omega_1$. The laser field is switched on shortly before the arrival of the photons, in order to avoid the atomic decay from the upper level $|F'=2,m_F'=+2\rangle_2$ into the states outside of the system during the PNSS storage. Similarly, the same magnetic field as in the sending system is applied to provide two-photon resonant Raman transition with linearly polarized cavity photons.

The complete conversion of PNSS into the superposition of ground Zeeman states of the receiving atom implies that the output quantum field from the second cavity is identically zero at all times. This amounts to formally setting $a_{2,\text{out}}(t)=0$ in the input-output relation $a_{2,\text{out}}(t)-a_{2,\text{in}}(t)=\sqrt{k}a_2(t)$ for the second cavity. Then, using $a_{2,\text{in}}(t)=a_{1,\text{out}}(t-\tau)$, we have
\begin{equation}
a_2(t) = -\frac{1}{\sqrt{k}}a_{1,\text{out}}(t-\tau),\label{eq:25}
\end{equation}
where $\tau$ represents the time for the photons to travel from one cavity to the other. We can eliminate $\tau$ according to the formalism of cascaded quantum systems \cite{18,19}, which include also our scheme, thereby $\tau$ can be omitted in what follows.

\subsection{Evidence of PNSS storage from conservation laws}
Here we derive the conservation law for the numbers of photons and atomic excitations. The effective interaction Hamiltonian for the receiving system has the form similar to Eq.\eqref{eq:3}
\begin{equation}
H=\hbar G_2 f_2^{1/2}(t) \biggl (a_2^{\dag}(t)(\varrho_{1,0}+\varrho_{0,-1})+(\varrho_{0,1}+\varrho_{-1,0})a_2(t)\biggr ),\label{eq:26}
\end{equation}
where $\varrho_{i,j} = |m_F=i\rangle_{22}\langle m_F=j |$ are the operators of the receiving atom, $G_2=g\Omega_2/\Delta$ and $f_2(t)$ is the second laser intensity shape. In the absence of relaxations, the equations for the photon number $a_2^{\dag}a_2$ and atomic population operators $\varrho_{i,i}(t)\equiv\varrho_i(t), i=0,-1,$ in the second cavity are obtained in the form
\begin{widetext}
\begin{subequations}
\begin{eqnarray}
& & \frac{d(a_2^{\dag}a_2)}{dt}(t) =-i k G_2\biggl [a_2^{\dag}(\varrho_{1,0}+\varrho_{0,-1}) -(\varrho_{0,1}+\varrho_{-1,0})a_2\biggr ]+k a_2^{\dag}a_2,\label{eq:27a}\\
& & \frac{d\varrho_0}{dt}(t)=i G_2\biggl(a_2^{\dag}(\varrho_{1,0}-\varrho_{0,-1})+(\varrho_{-1,0}-\varrho_{0,1})a_2\biggr),\label{eq:27b}\\
& & \frac{d\varrho_{-1}}{dt}(t)=i G_2(a_2^{\dag}\varrho_{0,-1}-\varrho_{-1,0}a_2)\label{eq:27c}.
\end{eqnarray}\label{eq:eq27}
\end{subequations}
\end{widetext}
By combining and integrating these equations and using Eq.\eqref{eq:25}, we derive the conservation law
\begin{equation}
N_0(t)+2N_{-1}(t)=n_{\text{out}}-\frac{1}{k}{\mathcal F}(t),\label{eq:28}
\end{equation}
where $N_{0(-1)}$ is the population of ground state Zeeman sublevel $m_F=0(-1)$ of the receiving atom, ${\mathcal F}(t)=\langle a^{\dag}_{1,\text{out}}(t)a_{1,\text{out}}(t)\rangle$ is the flux of outgoing photons from the sending cavity, and $n_{\text{out}}=\int \limits_{-\infty}^t{\mathcal F}(t')dt'$ is defined in Eq.\ref{eq:17}. From Eq.\eqref{eq:28} we recognize that $N_0(-\infty)=N_{-1}(-\infty)=0$, i.e. the receiving atom is initially in the state $|F=1,m_F=1\rangle_2$. For large times, taking into account that $N_1(\infty)=0$ and the photon number tends to $n_{\text{out}}(\infty)=1+|c_{-1}|^2$ following from Eq.\eqref{eq:20}, while $F(\infty)\rightarrow 0$, we find $N_{-1}(\infty)=|c_{-1}|^2$ and $N_0(\infty)=|c_0|^2$. This demonstrates that the sending and receiving atoms ultimately exchange initial states, thus implementing the complete transfer of the quantum information between the two network nodes. The Zeeman coherence for the receiving atom is next shown by analytical calculations of the population amplitudes with simultaneous determination of the second laser pulse shape.

\subsection{Coherent absorption of photons. Tuning of the laser pulse shape }
We expand the state of the receiving system as
\begin{equation}
|\Psi_2(t)\rangle = \biggl (\sum_{m_F=-1}^{1}\sum_{j=0}^{m_F+1}\gamma_{m_F,j}(t)|m_F\rangle_2|j\rangle\biggr ),\label{eq:29}
\end{equation}
where, similar to $\beta_{m_F,j}(t)$ in Eq.\eqref{eq:11}, $\gamma_{m_F,j}(t)$ are normalized amplitudes of the probabilities that at time $t$ the receiving atom is in the state $|m_F\rangle_2$ and the number of incoming photons is $j$. The level populations of the second atom are calculated by the formula
\begin{equation}
\langle\varrho_{m_F}(t)\rangle = \sum_{j=0}^{m_F+1}|\gamma_{m_F,j}(t)|^2.\label{eq:30}
\end{equation}
We find the coefficients $\gamma_{m_F,j}(t)$ from the Schr\"odinger equation for $|\Psi_2(t)\rangle$ with the use of Hamiltonian \eqref{eq:26}, where the operator $a_2(t)$ is replaced by $a_{1,\text{out}}(t)$ from Eq.\eqref{eq:25}. It should  be taken into account that the photonic state in the $\gamma_{m_F=0,j=1}$ term is the superposition of two one-photon states: $|j=1\rangle=\frac{1}{\sqrt{2}}(|1_{\Phi_1},0_{\Phi_2}\rangle+|0_{\Phi_1},1_{\Phi_2}\rangle)$. Then, by introducing new variables
\begin{subequations}
\begin{eqnarray}
& & \eta(t) = 2\frac{|G_2|}{\sqrt{k}}\int \limits_{-\infty}^tf_2^{1/2}(\tau)\Phi_1(\tau)d\tau,\label{eq:31a}\\
& & \zeta(t) = \frac{|G_2|}{\sqrt{k}}\int \limits_{-\infty}^tf_2^{1/2}(\tau)\biggl(\Phi_1(\tau)+\Phi_2(\tau)\biggr)d\tau\label{eq:31b}
\end{eqnarray}
\end{subequations}
and using Eq.\eqref{eq:23}, the equations for $\gamma_{m_F,j}(t)$ are derived in a simple form
\begin{subequations}
\begin{eqnarray}
\frac{d\gamma_{ 0,0}(t)}{d\eta} = \frac{i}{2}\gamma_{1,1}(t)e^{-i\varphi_2},\label{eq:32a}\\
\frac{d\gamma_{1,1}(t)}{d\eta} = \frac{i}{2}\gamma_{0,0}(t)e^{i\varphi_2}\label{eq:32b}
\end{eqnarray}\label{eq:32}
\end{subequations}
and
\begin{subequations}
\begin{eqnarray}
\frac{d\gamma_{-1,0}(t)}{d\zeta} = \frac{i}{\sqrt{2}}\gamma_{0,1}(t)e^{-i\varphi_2},\label{eq:33a}\\
\frac{d\gamma_{0,1}(t)}{d\zeta} = \frac{i}{\sqrt{2}}\biggl(\gamma_{1,2}(t)e^{-i\varphi_2}+\gamma_{-1,0}(t)e^{i\varphi_2}\biggr),\label{eq:33b}\\
\frac{d\gamma_{ 1,2}(t)}{d\zeta} = \frac{i}{\sqrt{2}}\gamma_{0,1}(t)e^{i\varphi_2}.\label{eq:33c}\nonumber
\end{eqnarray}\label{eq:33}
\end{subequations}
where $\varphi_2$ is the phase of $\Omega_2$ field. These equations describe the coherent excitation of two-level and three-level systems by "one- and two-photon" absorption, respectively, where the role of the "one-photon transition" in the first case is played by the first Raman transition $|F=1,m_F=1\rangle_2\rightarrow |F=1,m_F=0\rangle_2$ in Fig.\ref{fig:fig1} (right) with the effective pulse area $\eta(t)$. In the second case, the "two-photon excitation" is accomplished by the double Raman transition $|F=1,m_F=1\rangle_2\rightarrow |F=1,m_F=0\rangle_2\rightarrow |F=1,m_F=-1\rangle_2$ with total pulse area $\zeta(t)$.

\begin{figure}[t] \rotatebox{0}{\includegraphics*
[scale =0.7]{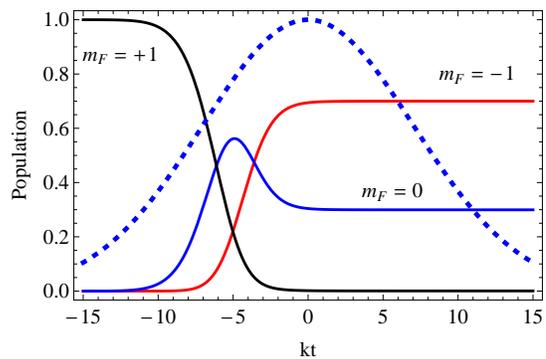}} \caption{(Color online) Evolution of the populations of ground Zeeman states $|m_F=-1\rangle_1$ (red), $|m_F=0\rangle_1$ (blue) and $|m_F=+1\rangle_1$ (black) of the second atom for the $\Omega_2 = \Omega_1$ Gaussian laser pulse of intensity profile $f_2(t)=e^{-(t/T_2)^2}$ with duration $T_2=1~\mu$s. The profiles of the laser fields in the first and second cavities are shown by the red and blue dotted lines, respectively. For the sending system, the same parameters as in Fig.\ref{fig:fig2} are used.
\label{fig:fig6}}
\end{figure}

Solving the Eqs.\eqref{eq:32} and \eqref{eq:33} subject to the initial conditions $\gamma_{1,2}(-\infty)=c_{-1}$ and $\gamma_{1,1}(-\infty)=c_0$ following from Eq.\eqref{eq:24}, we get for $\varphi_2=\pi/2$
\begin{equation}
\gamma_{0,0}(t) = c_0\sin[\eta(t)/2], \quad \gamma_{1,1}(t) = c_0\cos[\eta(t)/2]\label{eq:34}
\end{equation}
and
\begin{subequations}
\begin{eqnarray}
\gamma_{1,2}(t) = \frac{1}{2}c_{-1}\biggl [1+\cos\zeta(t)\biggr],\label{eq:35a}\\
\gamma_{0,1}(t)=\frac{1}{\sqrt{2}}c_{-1}\sin\zeta(t),\label{eq:35b}\\
\gamma_{-1,0}(t) = \frac{1}{2}c_{-1}\biggl [1-\cos\zeta(t)\biggr].\label{eq:35c}
\end{eqnarray}\label{eq:35}
\end{subequations}
It is evident that the complete mapping of photons onto the atomic ground states can be achieved by means of effective two-photon (Raman) $\pi$-pulses, which imposes the following  conditions on the pulse areas
\begin{equation}
\eta(\infty)=\zeta(\infty)=\pi,\label{eq:36}
\end{equation}
leading to $\gamma_{0,0}(\infty)=c_0$ and $\gamma_{-1,0}(\infty)=c_{-1}$ and zero for the remaining coefficients. As a result, the second atom is settled in the final state
\begin{equation}
|\Psi_{\text{2,fin}}\rangle=c_{-1} |-1\rangle_2 + c_0 |0\rangle_2,\label{eq:37}
\end{equation}
which coincides in form with the initial state of the sending atom in Eq.\eqref{eq:4}. It is obvious that the fidelity of the state $|\Psi_{\text{2,fin}}\rangle$ with $|\Psi_{\text{1,in}}\rangle$, which have commuting density matrices, is equal to unity. The populations of ground Zeeman sublevels of the second atom found from Eq.\eqref{eq:30} are displayed in Fig.\ref{fig:fig6}.

\begin{figure}[t] \rotatebox{0}{\includegraphics*
[scale =0.7]{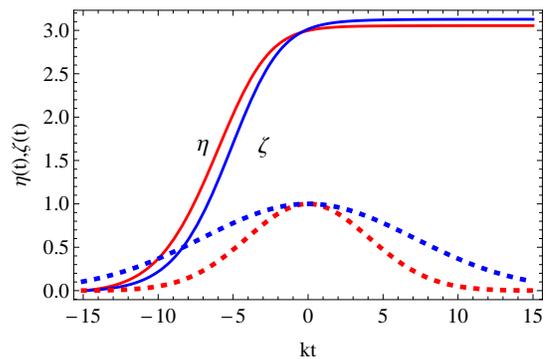}} \caption{(Color online) Variables $\eta(t)$(red) and $\zeta(t)$(blue) as functions of time for the same parameters as in Fig.\ref{fig:fig6}.
\label{fig:fig7}}
\end{figure}

The strength and temporal shape of $\Omega_2(t)$ pulse is fully determined by the conditions \eqref{eq:36}. For the used parameters, the latter are well fitted by Gaussian pulse with $\Omega_2=\Omega_1$ and duration $T_2\sim1~\mu$s. The time evolution of $\eta(t)$ and $\zeta(t)$ is shown in Fig.\ref{fig:fig7}. It should be noted that for the transfer of atomic quantum state, our scheme does not require time-symmetric photonic wave packets, in contrast to the scheme proposed in Ref.\cite{2}. The coherent absorption of arbitrarily shaped traveling photons in the second cavity is governed by the area of two-photon Raman pulses, where the time dependence of the second control field is appropriately adapted.

\subsection{Quantum state transfer via photonic qutrit}

The most important feature of multiphoton states is their ability to carry more information than a photonic qubit. One way to increase the information content of PNSS is to include into the superposition the states with more than two photons, which will allow to send information with increased capacity parameters. However, the current technologies do not support these statement, since along with the increase of the photon number, the losses in the communication channel increase exponentially. In contrary to this, our protocol is capable of producing a photonic qutrit without increasing the number of photons we use and, thus, preserving the robustness and efficiency of the PNSS qubit discussed in the next section. To this end, the atom is initially prepared in the superposition of the Zeeman sublevels $m_F=-1,0,+1$ in the sending node. Obviously, after interacting of the atom with the $\Omega_1(t)$ field, the generated photon state leaving the first cavity is analogous to the state \eqref{eq:13} and has the form
\begin{equation}
|\Psi_{\text {PNSS}}(t)\rangle=c_{-1}(t)|1_{\Phi_1},1_{\Phi_2}\rangle + c_0(t) |1_{\Phi_1}\rangle + c_{+1}(t)|0\rangle,\label{eq:13prime}
\end{equation}
where the coefficients $c_{-1},c_0,c_{+1}$ are the population amplitudes of the Zeeman sublevels  $m_F=-1,0,+1$, respectively, and all of them differ from $0$ at the initial time moment $t=-\infty$. Thus, the atom qutrit state is completely mapped into the photonic qutrit state. Using the same method, the state \eqref{eq:13prime} is transferred to the receiving atom, which is initially prepared at the Zeeman sublevel $m_F=+1$. Similar to the case of the qubit transfer, the second atom coherently absorbs the photons and  passes into the superposition state  of three ground-state Zeeman sublevels, which coincides in form with the initial state of the first atom.

\section{\protect\normalsize DISCUSSIONS}
At short communication distances, where the photon losses are negligible, the presented protocol is fully deterministic, relied on the high signal-to-noise ratio Eq.\eqref{eq:6} and enhanced Raman emission of linearly polarized cavity photons in the sending node and their controlled coherent absorption in the receiving system. An important advantage of the present scheme is the capability of multi-photon superposition states to transmit an incomparably greater information than is possible with a single-photon transfer. Imperfections of the scheme are manifested at large transmission distances caused by photon losses. Presently, the photon losses are comprehensively analyzed in \cite{Q} for two protocols of deterministic QST based on wave packet shaping \cite{2} and adiabatic passage \cite{P}. Remind that in the first case, the state transfer is realized by producing a time-symmetric photon wave packet inside a fiber and in the second one the QST is accomplished by performing the adiabatic passage with classical driving fields in a counterintuitive order. This analysis has shown that the state transfer by adiabatic passage can mitigate the effects of cavity losses, while the fiber transmission losses cannot be overcome using either of the two methods. Our approach does not fit into this paradigm, as we do not use the time-symmetric photon wave packets, and the control field in the receiving node is delayed with respect to the control field in the sending node. Nevertheless, our findings are analogous to those presented in \cite{Q}, demonstrating that the cavity losses can be strongly suppressed by making the signal-to-noise ratio sufficiently large, as shown in Sec.IIB, but due to fiber losses the QST is strictly limited by Eq.\eqref{eq:38} given below.

To show this we evaluate the efficiency of the QST by introducing the success probability as
\begin{equation}
P_{j,\text{trans}}=p_{j,\text{em}}\eta_{j,\text{trans}}p_{j,\text{abs}},\label{eq:38}
\end{equation}
with $j=1,2$ defined as the product of quantum-mechanical probabilities $p_{j,\text {em}}$ and $p_{j,\text {abs}}$ of $j$ photon emission in the sending and their absorption in the receiving nodes, respectively, by the transmission efficiency $\eta_{j,\text{trans}}=\text{exp}(-\frac{jL_0}{L_{\text{att}}})$, where $L_0$ is the distance between the nodes and $L_{\text{att}}$ is the communication channel attenuation length \cite{21}. It is apparent that the two-photon transmission efficiency is quadratically smaller compared to the one photon case: $\eta_{2,t}=\eta_{1,t}^2$. Today, the commercial fibers feature an attenuation of 2~dB/km at 800~nm \cite{21}, which corresponds to $L_{\text{att}}=2.2$~km. This means that in the present scheme with $p_{j,\text {em}}=p_{j,\text {abs}}\sim 1$ the success probability of PNSS transfer between the nodes connected by an optical fiber link of 60~m length, as in the experiment \cite{3}, is unity. At larger distances $L_0\sim L_{\text{att}}$, the efficiency of two-photon transfer decreases in our scheme by approximately an order of magnitude. However, it remains comparable with the efficiency of an ideal transfer of a single-photon polarization state in the scheme of Ref.\cite{3}, if one takes into account that in that scheme the probability $p_{1,\text {em}}$ of this state generation in the sending node is limited by the value of $25\%$. Nevertheless, as at these distances the fidelity of the state transfer process becomes dependent on its efficiency, a possible way to overcome this difficulty is to filter out successful events by a heralding signal. This is comparatively easy to implement for the creation of remote atom-atom entanglement, as evidenced by a large number of publications on this issue (see, for example, \cite{21}). As to the heralded QST, to our best knowledge, so far only one protocol has been proposed and realized for a heralded, high-fidelity mapping of the polarization state of a single photon onto an atomic qubit using Zeeman manifolds of two different atomic states \cite{E,kurz,F}. It is worth noting that the extended version of this scheme has been used in \cite{R} to entangle partner photons absorbed by ions in remote traps. At the same time, such an approach is not yet available in the case of multiphoton absorption, therefore the implementation of heralding remains challenging with our protocol. However, as suggested in \cite{3}, the successful storage of incoming photons in the second cavity can be verified by analyzing the Zeeman state populations of the second atom using the cavity-assisted fluorescence detection \cite{B,S,T}.

The superposition state architectures require that the phase acquired by photons in long fiber links must remain constant for times larger than the travel time. In this sense, the phase stabilization is a common problem and requires an identical solution in all protocols. In our scheme, two photons in PNSS, which are not correlated or entangled, have the same polarization, and suffer from the same losses and same polarization and velocity fluctuations, especially as they are efficiently outcoupled into a single-mode optical fiber with almost overlapped temporal profiles. Therefore, if only phase noise in optical fibers is considered, the two-photon state undergoes only an overall phase change, which can deteriorate the PNSS fidelity. However, in conventional conditions, the mean phase shift due to the fluctuations in fibers is about $0.1$~rad/km \cite{noise}, which is small enough to neglect the phase noise at distances where the photon losses start. This is also witnessed by the large values of the fidelity reaching 85\% for the transferred photonic polarization state in the experiment of Ref.\cite{3}. Moreover, as shown in \cite{U}, the width of the Gaussian distribution of the phase noise in the Mach-Zehnder interferometer is even lower, so that the phase remains constant for several tens of kilometers.

We conclude our discussion by returning to the well-studied question that the atomic motion in the cavity can be significantly affected by photon recoil from emission and absorption of even a single photon, which can no longer be ignored in the case of multiphoton processes. To remove this effect, for example in a photon emission, the atomic recoil $L_\text{rec}$ has to be made negligible with respect to the photon wavelength $\lambda$, where  $L_\text{rec}$ is calculated as the distance the atom travels during the time interval $T_\text{phot}$ between two time moments of the first and second photons emission: $L_\text{rec}=\hbar kT_\text{phot}/m$, where $m$ is the atomic mass, $k=2\pi/\lambda$. Since $T_\text{phot}$ is  reciprocal to the generation rate of photons in the cavity, $T_\text{phot}^{-1}\sim G_1=g\Omega_1/\Delta$, then from $L_\text{rec}\ll \lambda$ one finds $\omega_\text{rec}\ll G_1$, where $\omega_\text{rec}=\hbar k^2/2m$ is the recoil frequency associated with the kinetic energy transfer to the atom in a single photon scattering event. Usually, to localize the atom at the center of the cavity mode,  high-intensity laser beams along all three directions are used thus reducing the recoil frequency to about 4 kHz \cite{V}.  We assume that this is the case in our model, where $G_1\simeq 1.5$~MHz, so that the cavity emission does not change the motional state of the atom. This is true for the photon absorption in the second cavity as well.

 These observations show that to suppress the  environmental effects no additional efforts are needed in our scheme, while its advantages are obvious.

\section{\protect\normalsize CONCLUSION}
In conclusion, we have proposed and analyzed a robust and efficient scheme for deterministic transfer of quantum states between two remote cavity-trapped atoms by means of traveling photon-number superposition states.The quantum information encoded in ground Zeeman sublevels of the first atom is released in the photonic qubit or qutrit, depending on the number of superposed Zeeman states, via cavity-assisted Raman scattering and upon reaching the second atom is stored in the ground Zeeman states of the latter via coherent absorption of photons. We have demonstrated that our scheme allows reliable overall transfer of quantum information at the distances, where the photon losses are negligible, without using a quantum error correction.
At the same time, our scheme displays no less reliable protection against the propagation losses at large distances as compared to previous protocols with a single photon transfer.

These results promise also a successful distribution of entanglement between distant nodes in quantum networks. This study is currently under way.

\subsection*{Acknowledgments}

Authors thank to S. Gu\'erin and A. Kuhn for helpful discussions and comments. This research has been conducted in the scope of the International Associated Laboratory (CNRS-France $\&$ SCS-Armenia)  IRMAS. We acknowledge additional support from the Science Basic Foundation of the Government of the Republic of Armenia.

\end{document}